\documentclass[10pt,conference]{IEEEtran}
\IEEEoverridecommandlockouts

\usepackage{cite}
\usepackage[numbers]{natbib}
\usepackage{amsmath,amssymb,amsfonts}
\usepackage{algorithmic}
\usepackage{graphicx}
\usepackage{textcomp}
\usepackage[table,xcdraw]{xcolor}
\usepackage{graphicx}
\usepackage{booktabs} 
\usepackage{framed}
\usepackage{url}
\usepackage{enumitem}
\usepackage{hyperref}

\hypersetup{
    colorlinks,
    linkcolor={red!50!black},
    citecolor={blue!50!black},
    urlcolor={blue!80!black}
}
\def\BibTeX{{\rm B\kern-.05em{\sc i\kern-.025em b}\kern-.08em
    T\kern-.1667em\lower.7ex\hbox{E}\kern-.125emX}}
\definecolor{darkgreen}{rgb}{0.05,0.5,0.05}

\definecolor{darkred}{rgb}{0.5,0,0}

\newif\ifwescuts

\begin{document}


\title{From Organizations to Individuals: 
Psychoactive Substance Use By Professional Programmers}



%
\makeatletter
\newcommand{\linebreakand}{%
  \end{@IEEEauthorhalign}
  \hfill\mbox{}\par
  \mbox{}\hfill\begin{@IEEEauthorhalign}}

\author{\IEEEauthorblockN{Kaia Newman}
\IEEEauthorblockA{Computer Science and Engineering \\
\textit{University of Michigan}\\
Ann Arbor, Michigan, USA \\
kaian@umich.edu}
\and
\IEEEauthorblockN{Madeline Endres}
\IEEEauthorblockA{Computer Science and Engineering \\
\textit{University of Michigan}\\
Ann Arbor, Michigan, USA \\
endremad@umich.edu}
\linebreakand
\IEEEauthorblockN{Westley Weimer}
\IEEEauthorblockA{Computer Science and Engineering\\
\textit{University of Michigan}\\
Ann Arbor, Michigan, USA \\
weimerw@umich.edu}
\and
\IEEEauthorblockN{Brittany Johnson}
\IEEEauthorblockA{Department of Computer Science \\
\textit{George Mason University}\\
Fairfax, Virginia, USA \\
johnsonb@gmu.edu}}
\maketitle

\begin{abstract}
Psychoactive substances, which influence the brain to alter perceptions and moods,
have the potential to have positive and negative effects on critical software
engineering tasks. They are widely used in software, but that use is not well understood. 
We present the results of the first qualitative investigation of the experiences of, and challenges faced by, psychoactive substance users in professional software communities. 
We conduct a thematic analysis of hour-long interviews with 26 professional programmers
who use psychoactive substances at work. Our results provide insight into individual motivations
and impacts, including mental health and the relationships between various substances
and productivity. Our findings elaborate on socialization effects, including soft
skills, stigma, and remote work. The analysis also highlights implications for organizational
policy, including positive and negative impacts on recruitment and retention. 
By exploring individual usage motivations, social and cultural ramifications, and organizational policy, we demonstrate how substance use can permeate all levels of software development.
\end{abstract}


\section{Introduction}

Psychoactive substances, which influence the brain to alter
behaviors, perceptions and moods, are widespread throughout the world~\cite{Wadley2016HowPsychoactive}. They played a key role in the early history of computer science~\cite{Markoff2005WhatThe} and remain prevalent in software engineering to this day~\cite{Endres2022HashingIt}. They can have significant positive and negative effects on attributes associated with programming, such as focus~\cite{nida2018PrescriptionStimulants}, productivity~\cite{Kendall2016HackingThe} and creativity~\cite{Walsh2011DrugsThe, Jarosz2012UncorkingThe, Prochazkova2018ExploringThe}, but also
carry moral, social and legal concerns~\cite{SubstanceAbuseandMentalHealthServicesAdministration2020KeySubstance, Ransing2022CurrentState}. Despite the risks and benefits, 
current uses of psychoactive substances by software engineers are not well understood. 

We desire a foundational understanding of the experiences and challenges faced by
psychoactive substance users who are also software developers --- both to clarify the
landscape and dispel uncertainty, but also to provide actionable insights for decision
makers (e.g., for hiring, culture, and retention). 
We focus on substances such as 
prescription stimulants (e.g., Adderall), cannabis (e.g., marijuana), alcohol, 
mood disorder medications (e.g., Zoloft), and psychedelics (e.g., LSD). 
The legality of psychoactive drugs varies by locality and substance,
with usage rates increasing for programmers with the rise of work-from-home policies~\cite[Sec.~5.2]{Endres2022HashingIt}. At the same time, developers are
increasingly turning to prescription and recreational psychoactive drug use --- while working --- 
to alleviate health symptoms and improve productivity (Section~\ref{sec:AnalysisIndividual}). 

In this context, an effective investigation must (1) minimize preconceived biases
and expectations surrounding this morally- and legally-sensitive topic, 
(2) speak to a broad range of professionals across organizations and levels of
experience, (3) describe the lived experiences of users of psychoactive substances
instead of the opinions of others about them, and (4) admit useful conclusions at
multiple levels of modern software engineering. To the best of our knowledge, 
the closest related work either focuses on preconceived questions about one
substance (e.g.,~\cite{Endres2022HashingIt,Darshan2013AStudy}) or addresses broad groups
of developers, but not about substance use (e.g.,~\cite{Ford2022ATale}). 

We propose the first investigation of psychoactive substance users in modern
software development, using qualitative methods to draw rigorous conclusions 
from a collection of semi-structured interviews of personal experiences. While there are numerous studies that have used qualitative research methods as a way to gauge and report on a broader range of developers' experiences and opinions~\cite{Johnson2022ProgramlOnline, Chattopadhyay2021DevelopersWho, Huang2021LeavingMy, Ford2019HowRemote, Singh2019WomenParticipation, Kim2016TheEmerging}, this is the first qualitative study on psychoactive drug use in software development. Guided by archival data from a pre-survey of 799 programmers about general substance
use, we designed research questions focusing on five themes: 
health, self-regulation, social interaction, company culture, and company policy. 
We conducted hour-long interviews of 26 experienced software developers, 
placing special care on ethical recruitment and confidentiality,
with multiple independent annotators ultimately discovering over 170 relevant shared
concepts. 

We distill those thematic findings and structure our presentation of them through
three lenses: individual usage motivations, social and cultural ramifications, and
organizational policy. Our findings shed light on mental health, programming
enhancement, soft skills, remote work, drug policies, hiring and retention, and company 
culture --- and how they interact with the common, but not always spoken of, 
use of psychoactive substances. For example, at the organizational level, we find that for many substance users, 
anti-drug policies are unclear and ineffective; such policies
are viewed as indicative of corporate culture and may have a negative impact on hiring and retention. We also discuss a direct mapping, based on our sample,
between alcohol, cannabis and stimulants and positive and negative effects on
software engineering tasks (e.g., brainstorming vs. debugging vs. meetings, etc.). 

The contributions of this paper are:
\begin{enumerate}
    \item The first qualitative study of the personal experiences surrounding 
    psychoactive substance use by professional programmers ($n=26$), based on
    a thematic analysis 
    
    \item An explanation of individual substance use motivations and impacts, such as mental
    health considerations as well as substance use and productivity (including
    per-substance and per-task breakdowns)
    
    \item An explanation of socialization effects of substance use in software, such as the impact on soft skills, visible work use and stigma, and the effect of remote work
    
    \item An explanation of organizational policy implications, including
    policy clarity and effectiveness and impacts on recruitment and retention (both positive and negative) 

\end{enumerate}

\begin{framed}
    \noindent This paper discusses the use of substances that are illegal or may be dangerous in some contexts. The authors neither endorse nor condemn this behavior. Rather, the goal is to understand, present, and qualitatively analyze the lived experiences of psychoactive substance users working with software.
\end{framed}

\section{Background and Related Work}
\label{sec:background} 

We now cover related work concerning psychoactive substance use (in general and software contexts), and software development and mental health.

\vspace{2pt} 
\noindent\textbf{General Psychoactive Substance Use:} A \emph{psychoactive} (or psychotropic) substance influences the brain or nervous system and thus behavior, mood, perception and thought~\cite{WHO2022DrugsPsychoactive}. 
Alcohol, caffeine, cannabis, LSD, and nicotine are examples of such substances. Additionally, many medications prescribed for mood disorders (such as depression or anxiety) are psychoactive. Different substances have different cognitive impacts: for example, alcohol suppresses nervous system activity while stimulants increase alertness and focus via dopamine in the brain~\cite{nida2018PrescriptionStimulants}. Psychoactive drugs have a long history and have impacted multiple aspects of human culture, from recreation to war~\cite{Wadley2016HowPsychoactive}. Prevalence of use and legality vary by substance and area~\cite{SubstanceAbuseandMentalHealthServicesAdministration2020KeySubstance, Ransing2022CurrentState}. In this work we focus primarily on those substances that we find are likely to be used while programming (see Section~\ref{sec:preStudy}): cannabis, alcohol, prescription stimulants (e.g., Adderall, Ritalin), mood disorder medications (e.g., SSRIs, Wellbutrin), and psychedelics (e.g., LSD, microdosing). Notably, although officially psychoactive, we exclude caffeine due to its near-universal prevalence in software.

\vspace{2pt}
\noindent\textbf{Psychoactive Substance use in Software:} Psychedelics, such as LSD, have been associated with early software development~\cite{Markoff2005WhatThe}, with folk wisdom suggesting positive creativity benefits~\cite{Walsh2011DrugsThe}. Similar
creativity benefits have been suggested for alcohol~\cite{Jarosz2012UncorkingThe}, and micro-dosing~\cite{Prochazkova2018ExploringThe}. 

As for explicit research on the intersection of psychoactive substances and software, the limited prior work focuses on individual substances. Endres \emph{et al.} conducted a survey of cannabis use in programming, finding that a substantial proportion of their sample used cannabis while completing software tasks~\cite{Endres2022HashingIt}. Darshan \emph{et al.} linked alcohol and depression in a study of IT professionals in India~\cite{Darshan2013AStudy}. Popular media have also reported that Silicon Valley culture includes using stimulants or other ``smart drugs'' (e.g., nootropics) to increase productivity~\cite{Kendall2016HackingThe}. However, to the best of our knowledge, no formal work has studied the intersection of psychoactive substance use as a whole in software. 

\vspace{2pt}
\noindent\textbf{Mental Health and Software Development:}
The happiness of software developers has been correlated positively with their productivity and quality of their work~\cite{Graziotin2019HappinessAnd,Kaur2020OptimizingFor}, 
supporting what is commonly referred to as the ``happy-productive'' thesis~\cite{Zelenski2008TheHappyproductive,Taris2009WellbeingAnd}. 
Beyond this, the \emph{un}happiness of software developers has been identified to have dozens of potential negative outcomes~\cite{Graziotin2018WhatHappens}. It remains in a company's best interest to prioritize the happiness of its employees for the best results.
Unfortunately, a considerable stigma remains around discussing mental health 
or medication (e.g.,~\cite{Hipes2016TheStigma,Brown2011StandardizedMeasures}) or neurodivergence issues~\cite{Morris2015UnderstandingThe}, hindering constructive reform. 



\section{Pre-Study: Survey Results}

\label{sec:preStudy}


\noindent\textbf{Pre-Study Setup:} Endres \emph{et al.} surveyed 803 students and full-time programmers, finding that 35\% of their sample had programmed while using cannabis, that 18\% do so at least once per month, the primary motivation being to enhance certain software development skills (e.g., brainstorming) rather than for pain relief~\cite{Endres2022HashingIt}. That survey focused on cannabis alone rather than psychoactive substances in general (e.g., aspects like Adderall and ADHD were not directly included) and it primarily used pre-set questions rather than capturing distinct subjective experiences of users.
However, Endres \emph{et al.} provided additional archival data when requested, including data from 799 programmers who filled out a brief section related to the use of other (i.e., non-cannabis) substances. While it does not directly address our goal, 
this archival data provides a rich source of preliminary information to guide the construction of
semi-structured interview questions and motivate the
themes we explore. 

\noindent\textbf{Pre-Study Quantitative Results:} We first use this archival survey data to 
investigate the prevalence of substance use among developers. Participants were asked if they had used various psychoactive substances in the last year while completing software-related tasks. In that sample, 59\% (473/799) of participants reported using a mind-altering substance while completing a software-related task in the last year. 
Alcohol (25\%) and cannabis (24\%) were the most common. The next most common two were tobacco (6\%) and amphetamines (5\%, including both recreational and prescribed stimulants). 
Perhaps counter to common stereotypes of software developers~\cite{Markoff2005WhatThe,Prochazkova2018ExploringThe}, though the next most common, psychedelics use (including microdosing) was quite rare overall (2\%). 

To guide our qualitative investigation of usage patterns,
we also compare general use frequencies to those
in software contexts. Differences in this ratio between substances give an indication of
the ``normalization'' or ``uniqueness'' of a substance to software by its users. As a baseline, we find that 93\% percent of developers who used caffeine in the last year also report using it while doing software tasks. By contrast, only 50\% of alcohol users do so. These results align with our
intuition, and for drugs perceived as ``harder'' or
less socially acceptable, the percentages are even lower
(e.g., 30\% for cocaine and opioids, 22\% for hallucinogens). 
Intriguingly, other than caffeine, the substance with the highest transfer to software is amphetamines at 70\%, a transfer that is significantly higher than alcohol ($p < 0.01$). 
These differences --- both between substances, but also between work use and general use --- motivate our qualitative investigation of such usage patterns and motivations. 

\begin{table}[t]
    \centering
        \caption{Themes identified from a freeform question
        about programming and mind-altering substances in a pre-study.}
    \label{tab:prestudy-themes}

    \begin{tabular}{lp{0.70\columnwidth}}
        Theme           & Indicative Quote(s) \\ \midrule
        Self-Regulation & 
        ``Drugs are not to be used for everyday coding challenges, but I have had many coding breakthroughs while on low-risk mind-altering substances.'' \\

        Socialization &  
        ``I am more likely to want to collaborate when high \dots it allows me to have a better and more joyous time working with other people''\\ 
        & ``Communicating with people on mind-altering substances is a lot harder than people in their normal mental state'' \\ 
        
        Health   &  
        ``I take Adderall for ADHD treatment. It helps me focus on programming tasks by reducing the noise and intrusive thoughts in my brain.'' \\
        
        Culture &  ``Overall my company is pretty laissez-faire vis-a-vis cannabis. \dots I've often joked that if my company decided to start drug testing, they'd lose half the company.''\\

        & ``Microdosing is a truly revolting Silicon Valley trend'' \\ 
        
        Policy & 
        ``I\dots feel so terrified about my habit being discovered, and\dots whether that would impact my employment'' 
    \end{tabular}
\end{table}

\noindent\textbf{Pre-Study Qualitative Results}. Although the study by Endres \emph{et al.} was primarily quantitative~\cite{Endres2022HashingIt}, 
we were also able to examine prose results from a 
freeform question: ``Do you have any comments regarding programming and the use of mind-altering substances?''
Authors one, two, and four independently annotated participant responses to that question with their own codes and then met to discuss and agree upon the five most prominent themes. These themes are outlined in Table~\ref{tab:prestudy-themes}.

Although the data and analyses presented here were not previously reported by Endres \emph{et al.}~\cite{Endres2022HashingIt}, we claim no novelty regarding this data source and instead use it to guide the construction of our qualitative instruments. 
Informally, the pre-study gives confidence that we are pursuing the right questions.

\section{Main Study: Study Methodology}
\label{sec:study-design}
\label{sec:study-methodology} 


Guided by the quantitative and qualitative results from
the pre-study data, we developed five primary research
questions: 


\begin{description}
    \item [\textbf{RQ1}] What is the relationship between mental or physical health and the use of psychoactive substances in software working environments?
    \item [\textbf{RQ2}] What are the use and self-regulation patterns developers follow when using mind-altering substances for completing software tasks?
    \item [\textbf{RQ3}]  How does substance use impact in-person and remote social aspects of software working environments?
    \item [\textbf{RQ4}] How are different substances accepted or stigmatized in software workplaces?
    \item [\textbf{RQ5}] How do company drug policies impact developers who use psychoactive substances?

\end{description}

From these research questions we developed a semi-structured interview script. For the rest of the paper, we focus conducting and analyzing this interview with 26 software developers who use psychoactive substances while programming. 


\vspace{2pt}\noindent\textbf{Participant Recruitment:} We conducted 26 interviews using the aforementioned protocol (see Section~\ref{sec:popContext} for contextualization of our population). Participants had to be over the age of 18, work or have worked in a job that required developing software, and have significant first- or second-hand experience with using psychoactive substances at one of these jobs. 
We prioritized reaching participants with more years of experience in software, substantial experience with psychoactive substances, and diverse professional industries. Our decision to scope participants to those with direct experience of psychoactive substance use is a design choice as our research questions relate to software developers who use psychoactive substances. Additionally, our overall goal is not to collect a \emph{random} sample (indeed, there are ethical challenges to random samples of illegal activities), but to capture diverse opinions and experiences. 

We used a multi-pronged approach to recruit a spectrum of participants: physical posters, software-related mailing lists, word-of-mouth snowball sampling, and social media sites including Twitter and Reddit, as recommended for hard-to-reach populations~\cite{Sadler2010ResearchArticle}. More detail regarding our recruitment procedure (including specific subreddits used) is in our replication package. By the end of 26 interviews, we had reached saturation on our original research questions. It is standard practice in qualitative studies to stop collection when reaching or approaching saturation~\cite{Kim2016TheEmerging}.

\vspace{2pt}
\noindent\textbf{Interview Protocol: } Each interview was \emph{semi-structured} and lasted one hour, a design that both permits us to answer research questions and allows for unexpected themes to arise naturally in conversation~\cite{McIntosh2015SituatingAnd,Whiting2008SemistructuredInterviews, Seaman2008QualitativeMethods}. Generally, two researchers attended each interview: one asked most of the questions while the other took notes. The audio was recorded and later manually transcribed into text by the first two authors. 

Our full interview script is available in our replication package\footnote{Replication materials are available on Zenodo~\cite{Newman2023MaterialsFor} or on GitHub at \url{https://github.com/CelloCorgi/ICSE2023_Psychoactive}. We note this package does not include interview transcripts to protect the privacy of our participants.} At a high level, the interview started by learning more about the interviewee's professional programming experiences and to verify they were eligible to participate. The interview progressed through a series of overall topics connected to our research questions: 1) Basic experience with psychoactive substances in software, 2) Mental and physical health (e.g., ``Do you use substances to combat stress from work?''), 3) Social Impacts and Policy (e.g., ``Do others at work know that you use substances during software tasks?''), 4) Self Regulation (e.g., ``For which software tasks would you use a substance?''), and 5) Hypotheticals (e.g., ``If you could change anything about software drug culture, what would it be?'').

\if False
Our overall goal in this qualitative research is not to collect a \emph{random} sample (indeed, there are ethical challenges to random samples of illegal activities), but a sample that captures diverse opinions and experiences regarding our research questions. To be eligible, participants had to be over the age of 18, work or have worked in a job that required developing software, and have significant first- or second-hand experience using psychoactive substances at one of these jobs. We prioritized reaching out to participants with more years of experience in software, substantial experience with psychoactive substances, and diverse professional industries. 


We used a multi-pronged approach to recruit a spectrum of participants: physical posters, software-related mailing lists, word-of-mouth snowball sampling, and posting on social media sites including Twitter and multiple subreddit forums on Reddit, as recommended 
for hard-to-reach populations~\cite{Sadler2010Snowball}. 

The Reddit social media site required the most care to protect potentially-vulnerable populations. Reddit is a popular social news site containing many sub-forums (``subreddits'') with thriving discussions on  different software industries and psychoactive substances~\cite{Eghtesadi2020Social}. 
We contacted the moderators, 
followed any posted Study Guidelines 
(\href{https://www.reddit.com/r/microdosing/comments/ut5zgg/rmicrodosing_research_contribution_opportunities/subreddit}{example})
and formatting
requirements, 
validated our credentials and emails, 
and posted via an anonymous account with
a ``Contribute to Research'' flag to show community 
members that the post was approved. 
Two indicative examples of subreddits used are \texttt{r/microdosing} and \texttt{r/adderall}.

We first contact the moderators of a subreddit to see if they would be willing to host a message about participant recruitment. Reddit posts were made from an anonymous account related to the researchers’ group.  The \texttt{r/microdosing} has a Research Library link with a list of studies previously accepted to be posted and a list of Study Guidelines. Our proposed post was formatted and lengthened to model previously accepted studies, was given a “Contribute to Research” flag to show community members the post was approved by moderators, and our Reddit username was validated through an email authentication process.
Clarifications made by the moderators of r/microdosing reflected the privacy and legitimacy concerns of the community as a whole, especially since [many?] psychedelics are federally illegal. By contrast, the moderators of r/adderall said to “just make sure the post itself can be read very quick with nothing extra - this is what works best on r/adderall.”
    \begin{itemize}
        \item Intro: Procedure for contacting vulnerable population 
    \item Outro: [ conclusion / organization - there are multiple communities w/ differences …]
    \end{itemize}

The rest of the subreddits we reached out to and the subsequent posts on those approved are contained in our supplementary materials. FIXME - deal with supplementary materials

Each recruitment method included a link to a pre-screening form.
This included questions about eligibility, willingness, 
location and demographics. 
However, factors such as race, name, location or gender were not used to determine invitations to participate. Instead, they were used to distinguish
dishonest actors (informally, scammers who would duplicate
false identities for the interview compensation). 
We still reached out to participants who did not enter real names or physical addresses, if they provided context for nondisclosure. Candidates who passed
pre-screening were provided with
interview logistics, a Google appointment
scheduler, and a consent document. 
Our approach helped select a population of participants that had a wide range of experiences with substance use and software while also minimizing fraud.

Once we decided to contact someone, we used the email address they provided in the pre-screening form to send them a prewritten, standard invitation describing the basic structure of the interview, that it would be held on Zoom, and a Google appointment scheduler. Consent information was attached via a PDF document. When participants selected a one hour slot on our appointment scheduler, they could not see slots that had been taken by others, just that those times were not available anymore.

Our qualitative study approach requires balancing participant privacy against verifying identities during recruitment. In addition, we intentionally neither used students to approximate the experiences of
real developers nor focused our study on employees at one specific company. 
Our approach yielded a population of real-world software workers who used a variety of substances, both recreational and medicinal, prescribed and non-prescribed, from a variety of industries.

\begin{itemize}
\item We did not just use students to approximate the experiences of real developers
\item We did not do this study on employees at one specific company because we needed to be mindful of privacy and that’s too small of a scope for us wanting to report on how software devs generally use substances to set this field up for further research because it’s \textbf{very novel}
\end{itemize}
\fi 




\if False
The goals of our interview procedure were to maximize the amount of relevant information from each participant, to maximize participant freedom to bring up 
topics we had not considered a priority, and to be consistent in the prompts
each participant responded to. 
We balanced adaptability (e.g., to encourage elaboration on unique situations) with 
structure (to compare between different experiences and opinions on the same topic). 
Because of this, we chose to use a \emph{semi-structured interview} (SSI) format~\cite{McIntosh2015Semi,Whiting2008Semi}. 
While a structured interview has a fixed set of questions asked with no deviations and an unstructured interview has no prearranged questions, a semi-structured interview has a sample set of prearranged questions that can allow for further elaboration, as well as alternative questions depending on the interviewee responses. Semi-structured interviews enhance
participant trust and support discovering results not considered a priori. 

Our interview structure had rote beginning and ending procedures that were repeated verbatim to each participant. At the beginning we included a debriefing portion to obtain consent and remind participants of what was expected throughout the interview. We recorded interviews after
obtaining verbal consent. After the interview questions were over, we stopped recording and used a script to obtain payment information (etc.). We employed a bank of 50 questions that were categorized by research aim: basic experience, clarifying language, mental and physical health, social impacts, enhancement vs. alleviation, and ending questions and hypotheticals.
\fi

\vspace{2pt}
\noindent\textbf{Study Design Ethical Considerations -- Data Availability:} 
We highlight three ethical issues in our study design:
recruiting, confidentiality and informed consent.
For \emph{recruiting}, we did not require official 
emails or company names, and allowed interviews
with the camera off (after pre-screening with it on to verify identity). 
We employed a particularly high standard of \emph{confidentiality}, including not releasing the interview text
(which may contain identifying information admitting retaliation) 
without a data sharing agreement (e.g., via additional
IRB certification, etc.). Finally, 
we obtained IRB permission to waive written \emph{informed consent}
(a paper trail linking the participant's real name to the research) in favor of oral informed consent. 
These steps are pragmatically necessary (e.g., so people
feel comfortable speaking truths), but, more importantly, are
ethically necessary (to protect participants). 

\vspace{2pt}
\noindent\textbf{Data Analysis Methodology:} To analyze the interviews, we used a two-pass approach for tagging and annotating our data with codes: a first pass using a dynamic initial code book and a second pass with the finalized code book, both using the qualitative analysis tool ATLAS.ti. 
Our initial and final code books (and supporting quotes) are in our replication package.

In the first pass, each interview was coded using an initial code book derived from the interview outline. Two authors independently coded each transcript using this code book while also noting additional codes or themes encountered. Then, all authors met to merge findings. Using the union of codes and themes, three authors worked together to facilitate consensus on the emerging themes in the data and build a second, more robust code book. Independent codings of the same transcript were merged through group discussion rather than an automated process. The final code book consists of over 170 distinct codes organized into 10 groups. 

We do not formally calculate annotator inter-rater agreement of this first pass (see McDonald \emph{et al.}~\cite[Section 5.1]{McDonald2019ReliabilityAnd} for a discussion of reasons for and against calculating inter-rater agreement for qualitative studies). However, we retroactively examined 352 quotations from our transcripts to get an approximate understanding of inter-rater agreement in our study. In this sub-sample, 78\% matched a quote by the other coder (same or overlapping quotes with the same codes). Of the remaining quotes, 23\% were identified by both annotators but had at least one different code. Only 16\% of quotations were unique to one annotator, indicating relatively high consensus. 

In a second pass, an author annotated each interview with the complete code book. When possible, this author was \textit{not} one of the authors responsible for the first pass. Thus, at least three authors analyzed the majority of the interviews. 

After coding and tagging the data, two of the authors independently organized the codes into larger themes surrounding substance use in software. With a help of a third author, these themes were merged and organized into the three main levels that we present in this paper: individual motivations for substance use (Section~\ref{sec:AnalysisIndividual}), socialization effects of substance use (Section~\ref{sec:AnalysisSocial}), and organizational impacts (Section~\ref{sec:AnalysisOrganization}).

\section{Population Contextualization}

\label{sec:popContext}
We now describe the demographics and programming experiences, and psychoactive substances used by our population to better contextualize and scope our findings. 

\vspace{2pt}
\noindent\textbf{Demographics:} Of our 26 participants, 16 are men, 9 are woman, and 1 is non-binary. Participants range in age from 20 to 44, with an average of 30. As for location, 19 are based in the United States, two in India, and one each in Australia, Bangladesh, Mexico, the Philippines, and South Africa. An additional two participants moved to the US mid-career (from Israel and India), leaving 17/26 participants with US-exclusive experiences. We discuss the US-bias in our population more in Section~\ref{sec:threats}.

\vspace{2pt}
\noindent\textbf{Programming Experience:} All participants in our study have 1--20+ years of professional programming experience, with most having three or more years. 
The bulk (21/26) are current or former full-time software developers. The remaining five include one who owns and works at their own tech-related consulting company, one programming freelancer and student, one data analyst, and two computing-related Ph.D. students (both in the final stages of their degrees). 
Our participants work at companies with a wide array of sizes and industries: 5 at large software companies (e.g., FAANG, etc.), 9 at medium-sized software companies (1,000--3,000 employees), and 5 at smaller startups with under 500 employees. Additionally, 5 work in the financial sector, 2 work in the Health Care Sector, and 2 participants work at government contractors. Our replication package details participant experiences.

\begin{table}[t]
\centering
\caption{Number of participants in our sample that have used each psychoactive substance while
\label{tab:substancesUsed}
completing software tasks}
\begin{tabular}{@{}lr@{}}
\textbf{Substance}                                 & \textbf{\# of Participants} \\ \midrule
Prescription Stimulants (e.g., Adderall, Ritalin)  & 21                              \\
Cannabis (e.g., Marijuana, Weed)                   & 14                              \\
Alcohol (e.g., Beer, Wine)                         & 13                              \\
Mood Disorder Medication (e.g., SSRIs, Wellbutrin) & 11   \\                    
Psychedelics (e.g., LSD, Psilocybin)               & 7                               \\
Tobacco (e.g., Cigarettes, Vapes)                  & 7                               \\
Cocaine                                            & 3                               \\
Benzodiazepines (e.g., Xanax)                      & 2                               \\
Opiates (e.g., Codeine)                            & 2                               \\ 
\end{tabular}
\end{table}

\section{Findings Overview and Substances Used}

The findings from our interviews provide insights regarding psychoactive substance use in software development. Table~\ref{tab:substancesUsed} outlines the various substances two or more participants have used while developing software. 

The high number of Cannabis and Alcohol users in our sample correspond to the trends in our pre-study (Section~\ref{sec:preStudy}), as do the smaller proportions of psychedelics and tobacco. In contrast, however, the most common substance in our sample was prescription stimulants, which help increase focus and executive functioning. Commonly prescribed for various health conditions (see Section~\ref{subsec:AnalysisMentalHealth}), they can also be used recreationally. We discuss this discrepancy further in Section~\ref{sec:threats}. 

Our findings suggest that use of these psychoactive substances has the potential to impact all levels of a developer's work life from the individual to the organization. In this section, we outline results across three levels of developer experiences: the Individual (Section~\ref{sec:AnalysisIndividual}), Social (Section~\ref{sec:AnalysisSocial}), 
and Organizational (Section~\ref{sec:AnalysisOrganization}) impacts of psychoactive substance use in software. Commonalities, such
as Productivity and Work from Home, are discussed in each section. Additionally, we follow each included quote with a bracket noting the participant number (see our replication package for demographic details for each participant) and country location at time of the interview to better contextualize our findings.  

\section{Individual Motivations and Impacts}
\label{sec:AnalysisIndividual}

We start by focusing on the individual developer and their psychoactive substance use. We analyze reasons for, the personal effects of, and how and why programmers self-regulate their psychoactive substance use. Findings in this section address \textbf{RQ1} and individual aspects of \textbf{RQ2} (see Section~\ref{sec:study-methodology}).

At a high level, we observed two primary motivations for psychoactive substance use while programming: to help alleviate symptoms from mental health conditions (e.g., depression or ADHD), or to enhance programming abilities (e.g., creativity or productivity). In contrast, we do not observe physical health or addiction issues to be primary motivators.

While our findings suggest that, in practice, programming enhancement and mental health symptom alleviation may go hand in hand for many developers (especially in the use of prescribed substances such as stimulants), for clarity, we present substance use for mental health and for programming enhancement separately. We then conclude by detailing which substances our participants use for which software tasks. 

\subsection{Substance Use for Mental Health}
\label{subsec:AnalysisMentalHealth}
In our sample, mental health is a primary driver of psychoactive substance use when developing software: twenty of our participants reported using at least one substance prescribed by a psychiatric professional. 

\vspace{2pt}
\noindent\textbf{Mental Health: What Conditions?} The most common diagnosis in our sample is for \emph{Attention-Deficit/Hyperactivity Disorder} (ADHD), a neurodevelopmental condition marked by patterns of inattention (e.g., difficulty focusing), hyperactivity, and/or impulsivity. 15 participants have or are suspected by a psychiatrist to have ADHD. ADHD is often treated with prescription stimulants (e.g., Adderall) which improve focus and attention by increasing the amount of dopamine in the brain. In our sample, most  participants using stimulants for ADHD were diagnosed with ADHD in part or in whole due to symptoms present during their software work, and generally cite positive impacts of stimulants on software work. Supporting neurodivergent programmers, such as those with ADHD or Autism Spectrum Disorder, is increasingly important and visible in software engineering~\cite{Morris2015UnderstandingThe}. We consider the connection between ADHD and stimulant medication in greater detail later in this section. 

Aside from ADHD medications, we also spoke with 11  participants who are prescribed mood disorder medications (e.g., SSRIs, Wellbutrin, etc.) for depression or anxiety. 
In contrast to stimulants, participants were more dismissive of the effects of these substance on software. They often described mood disorder medications as not impacting work directly, but instead as removing obstacles making it difficult, if not impossible, to work. For example, one participant with diagnosed depression said: \textit{``I never related \dots  antidepressants and software work, because for me antidepressants and just overall feelings [are] not related to coding at all. Of course, it affects my software work. If I do have depression, I cannot work''} 
[P3, US].

\vspace{2pt}
\noindent\textbf{Symptoms at Work:} Many of our participants reported that their work triggers their mental health disorders. In some cases, symptoms at work contributed (in whole or in part) to diagnosis and treatment. This pattern was particularly common in participants diagnosed with ADHD. Of the 15 participants prescribed medication for ADHD, 11 were diagnosed in adulthood or seeking diagnosis while working professionally in software development. For these participants, the impact of prescription stimulants on their software work is almost uniformly positive: \textit{``Oh, it's been almost life-changing. It’s been wonderful. I'm way more focused obviously, but I'm also way more productive. It's a lot. Even when there's distractions at hand, I am able to manage those distractions better. I'm able to focus on my tasks better. I'm able to complete things in a faster time-frame than I could before. And it makes me almost want to start working each day''} [P5, US].

Because ADHD is always present from childhood, and most diagnoses are in adolescence~\cite{Holland2019RelativeAge}, it is surprising that most of our participants were diagnosed in adulthood. 
While ADHD diagnoses are rising overall, this discrepancy points to something software-specific. 
We hypothesize that participants may have previously experienced mild symptoms that had not interfered with daily life, but that the rigors of modern software development (and company culture and organization) made the symptoms impossible to ignore or mitigate non-medically.

\if False

\label{subsec:AnalysisMentalHealth}

\emph{Attention-Deficit/Hyperactivity Disorder} (ADHD) 

Although a comparable number (11/26) of participants noted that they take SSRIs (e.g., to
alleviate mood disorders such as anxiety or depression), many were more dismissive of their 
effects on their work lives. Many participants described SSRIs not as impacting their work directly but instead as removing obstacles that were making it impossible for them to work at all. 
One participant who had to take a month of leave due to depression said, ``I never related \dots  antidepressants and software work, because for me antidepressants and just overall feelings [are] not related to coding at all. Of course, it affects my software work. If I do have depression, I 
cannot work.''


Participants who reported a greater effects of SSRIs on work performance usually had specific anxiety or depression about work. One participant said, \textit{``I wouldn't feel anxious if I wasn't working usually, but when I would start working, I would just get anxious and I would choke up. That's something that happened to me a lot in that period of time. It doesn't happen anymore. I think the SSRIs helped with that.''} 

\vspace{2pt}
\noindent\textbf{Symptoms at work:} We find that many participants deal with symptoms from
mental conditions at work, those symptoms are often triggered by work, and developers
often alleviate those symptoms using psychoactive substances. 
Many (FIXME-KAIA/26) reported not being force themselves to work while depressed or
otherwise having their work quality impacted. One common way in which such symptoms
might be triggered at work is FIXME-KAIA (FIXME-KAIA/26). 

The process of responding to such symptoms, including by seeking a formal diagnosis, varied widely between participants. 
FIXME/26 were prescribed substances to alleviate symptoms directly, while others were subjected
to more intense (e.g., up to multiple days) scrutiny. Of the 15 participants that were diagnosed with ADHD or prescribed medication to alleviate ADHD symptoms, 11 were diagnosed in adulthood, mostly seeking this diagnosis or being prescribed medication over the course of their work in software. Because ADHD is a disorder that is present from childhood in all cases, and most diagnoses are in adolescence~\cite{Holland2019RelativeAge}, it is intriguing 
and surprising that most of our participants with ADHD were diagnosed in adulthood. 
While ADHD is better understood now, and thus diagnoses are rising overall, the striking
rate discrepancy in our data points to something software specific. 
We hypothesize that participants experienced symptoms levels that had not previously
interfered with activities of daily living (or could be alleviated with personal, non-medical 
compensations) but that the rigors of modern software development (and company culture and
organization) made the symptoms impossible to ignore or mitigate non-medically. This
caused them to to seek out a formal diagnosis or prescription medication. 

\vspace{2pt}
\noindent\textbf{Substance use to alleviate symptoms:}
FIXME is the most frequent substance used to alleviate mental health symptoms (FIXME/26), 
but FIXME is also common (FIXME/26). While some (FIXME/26) participants do report
using substances recreationally to treat mental health concerns (e.g., smoking marijuana
to calm down and address anxiety), participants generally favored prescription medications
while programming. However, we do observe a social tension between participants
who are \emph{prescribed} the use of stimulants (e.g., a participant diagnosed with ADHD
and prescribed Adderall) vs. those who use stimulants \emph{recreationally} (e.g., a
participant who chooses to use Ritalin to help with focus and gain a performance advantage). 
One participant noted,
``This sounds hypocritical, but I think it's more acceptable that people who use the substance [to] fix an actual problem instead of using them to improve their performance beyond what they are normally capable of. \dots So a person who has ADHD and they get prescribed Ritalin, and they take it everyday \dots That's perfectly fine. [But using] to outperform their peers and get a promotion then, and they don't have ADHD, I think it's unfair.'' 
Another summarized, 
``I have ADHD and I find it annoying when people take substances to get an advantage over others.''
This potential source of 
conflict is particularly relevant as many companies consider performance incentive
strategies (cf.~\cite{Roberts2006Understanding}) or have cultures that emphasize productivity. 
\fi

\begin{framed}
\noindent ADHD is the primary mental health diagnosis driving substance use in our sample. The majority of those diagnosed cited their seeking diagnosis was at least in part due to symptoms present in their software work. Stimulants prescribed as a result of diagnosis are viewed to have a positive impact on their software development.
\end{framed}

\noindent\subsection{Substance Use for Programming Enhancement}
\label{subsec:AnalysisSWWorkProcess}

\begin{table}[t]
    \centering
    \caption{Positive ($\uparrow$), Negative ($\downarrow$) and Neutral (--) assessments of
        \label{tab:drugEffectsOnTasks}
    substances on tasks or attributes of the development process.
    } 
\newcommand{\posE}[0]{\cellcolor[HTML]{BCEDA7} $\uparrow$~} 
\newcommand{\negE}[0]{\cellcolor[HTML]{FFFFC7} $\downarrow$~} 
    \begin{tabular}{lrrrr}
\textbf{Task}                        & $n$ &  \textbf{Alcohol}      & \textbf{Cannabis}      & \textbf{Stimulants} \\
\midrule 
Brainstorming       & 16 &  n/a          & \posE (86\%)      & -- (50\%) \\
Coding \& Testing   & 14 &  n/a          & -- (50\%)         & \textbf{\posE (91\%)} \\
Data Analysis       & 11 &  n/a          & \negE (33\%)      & \posE (88\%) \\
Debugging           & 21 &  \posE (100\%)& \negE (25\%)      & \textbf{\posE (93\%)} \\ 
Design              & 15 &  n/a          & \negE (40\%)      & -- (58\%) \\
Documentation       & 14 &  n/a          & \negE (25\%)      & \textbf{\posE (70\%)} \\
Meetings            & 12 &  n/a          & \negE (38\%)      & -- (50\%) \\ 
Requirements Elicit.& 5  &  n/a          & n/a               & n/a \\
\midrule 
\textbf{Individual Attributes}  &    &               &                   & \\
\midrule 
Creativity          & 13 &  \posE (100\%)& \posE (100\%)     & \negE(25\%) \\
Enjoyment of Work   & 12 &  \posE (100\%)& \posE (89\%)      & \posE(80\%) \\
Focus \& Productivity&25 & \posE (60\%) & \negE (33\%)      & \textbf{\posE(95\%)}\\
Quality of Work     & 12 & \negE (33\%) & \negE (40\%)      & n/a\\
\midrule 
\textbf{Social Attributes}  &    &               &                   & \\
\midrule 
``Soft'' Skills     & 18 &  n/a          & \negE (40\%)      & \textbf{\posE(80\%)}\\

    \end{tabular}
    
    $n$ counts participants that mention the task in conjunction
    with a substance. Percentages give the fraction of positive
    and negative mentions that were positive. Combinations with fewer
    than four mentions are not analyzed (``n/a''); positive or negative
    outcomes with 10 or more mentions are bolded. 
\end{table}

Aside from mental health support, participants also use psychoactive substance for programming ability enhancement. As described by one participant, \textit{``I want to be better. You know? I see myself like an athlete. And for me, [psychoactive substances] are like performance enhancement drugs''} [P13, US].

To contextualize the landscape of psychoactive substance use in software, we examine \textit{which attributes} of programming users seek to enhance. We identified four common attributes that are impacted through substance use: Creativity, Enjoyment, Work Quality, and Focus/Productivity. 
Table~\ref{tab:drugEffectsOnTasks} contains an overview of our results on how each of these attributes interacts with alcohol, cannabis, or stimulant medications, the three most common substances used by our population.

\noindent\textbf{Substance Use and Productivity:} All but one participant mentioned substance use for productivity enhancement. The substance most associated with productivity was stimulants. We did not explicitly ask about productivity. However, all 21 stimulant users stated that use increases their focus and productivity on certain software tasks. In fact, they often did so multiple times: increasing productivity and stimulants appeared together a total of 96 times in our data.

While stimulant use was commonly cited as having a positive effect on productivity, also common was mention of perceived \textit{decrease} in productivity with cannabis use.
This decrease in productivity was typically cited as a reason not to use cannabis for any given software task. Overall, however, our results indicate that enhancing productivity is a primary motivation for psychoactive substance use in software, a link that may speak to deeper threads of productivity culture in software culture as a whole. Touching on this culture, one stimulant user who works at a large FAANG-category company 
explained \textit{``I think it is really generous of them to offer the mental health benefits \dots and kind of say \dots `Hey, you should take care of your mental health.' \dots [However,] the way the performance reviews work is, at the end of the year \dots you basically have to catalog everything you did [and] show that you did all of it. And it can be pretty intense and it's pretty common for people to say things like, `Oh, I feel like I can’t take [personal time off] because then I’ll get behind on work''} [P10, US].

\noindent\textbf{Other Attributes:} After productivity, creativity was most common (45 mentions):
participants made numerous mentions of increased creativity and substance use, primarily with cannabis or psychedelics. Next was enjoyment (42 mentions), usually with cannabis or alcohol. Work quality was mentioned the least (34 times). 

\begin{framed}
\noindent When using substances for programming enhancement, increasing productivity is the most common goal, especially when it comes to using stimulants. Increasing creativity, work quality, or work enjoyment are cited less commonly, though when they are, it is usually in the context of alcohol, cannabis, or psychedelics. 
\end{framed}

\subsection{Self-Regulation During Software Tasks}
\label{subsec:AnalysisSelfRegulation}

Our findings suggest that developers may self-regulate substance use by software task. As seen in Table~\ref{tab:drugEffectsOnTasks}, participants associate tasks that require focus (e.g., debugging) with stimulants and tasks that require creativity (e.g., brainstorming) with cannabis or psychedelics. This implies \textit{a}) many developers are
deliberate about \textit{when} in the software process they use various psychoactive substances, treating it like a \textit{tool}, and \textit{b}) policies that ban certain substances in all cases may preemptively remove that tool from a developer's toolbox. We consider these implications in greater detail in the discussion. 

\vspace{2pt}
\noindent\textbf{Debugging:} The software task most mentioned with psychoactive substance use was debugging. Seen as a focus-intense and detail-oriented task, 14 participants reported that stimulants in particular are helpful for debugging. As stated by one such participant, \textit{``like when I'm debugging, sometimes you're all over the place, right? So there's a lot of things to keep in your mind at once \dots And I find that it's a lot harder for me to do that without Adderall''} [P26, US].


\vspace{2pt}
\noindent\textbf{Brainstorming:} Participants find brainstorming to be enhanced primarily by cannabis or psychedelics. Both are viewed as ways to see things from a new perspective. For example, when faced with solving a problem that left several other senior engineers stumped, one participant discussed using MDMA (a hallucinogenic stimulant) to help brainstorm the solution. 
In this participant's opinion, \textit{``there were huge boosts in creativity. I think it helped me big time in being in somewhat more of a na\"ive state and let go of everything that people had told me about the problem and kind of look at it from like my own lens \dots I have the personal opinion that responsible usage of [MDMA] is actually in the best interests of the company. I mean, I solved the very, very hard problem that four other engineers had failed. And I used these drugs''} [P21, Australia].
\begin{framed}
\noindent Developers choose to use different substances for different software tasks (e.g, stimulants for debugging, but cannabis for brainstorming), evidence that developers self-regulate their substance use, informally using it analogously to other development tools. 
\end{framed}

\section{Socialization Effects}
\label{sec:AnalysisSocial} 

While psychoactive substance use motivations are personal, use impacts can spill over into a developer's social network, affecting relations with co-workers and managers. In this section, we analyze how psychoactive substance use impacts socialization and interpersonal relations in software workplaces. We focus on choosing when to use psychoactive substances (``soft'' skills) and the stigma and visibility of substance use in software work environments. Findings in this section address social aspects of \textbf{RQ2}, \textbf{RQ3}, and \textbf{RQ4} (see Section~\ref{sec:study-methodology}).

\vspace{2pt}
\subsection{Social Impacts on Drug Self-Regulation}
\vspace{2pt}
\noindent\textbf{Substances and ``Soft'' Skills:}
Beyond individual technical skills, professional software development also requires significant interpersonal communication and interaction (``Soft'' Skills)~\cite{Matturro2019ASystematic,Ahmed2015SoftSkills}. Developers using substances for software tasks that require such soft skills often consider both their own performance and also the impact that use has on co-workers. 

Participants generally perceived soft skills to be \emph{improved} by substance use. Eight out of 10 stimulant users in our sample mentioning soft skills found that they helped with staying engaged and active in communication with other developers. As one stated,
during long meetings \textit{``[stimulants] allow me to become more engaged in what’s going on instead of my mind drifting into
\dots whatever I find more interesting,
which is basically everything else at that point''} [P16, US]. Mood disorder medications were also beneficial for communication, helping developers lower anxiety around presentations or stand-up meetings. For cannabis, however, we observed more mixed opinions: two out of five cannabis users perceived a positive impact on soft skills (vs. three negative). On the positive side, for one participant, cannabis lowers his anxiety when he \textit{``need[s] to do something like writing an email or talk to somebody about
some urgent topic, it’s easier for [him] to smoke and do it than do it sober''} [P3, US].
In contrast, when asked if her cannabis use differed between meetings and solo coding, another participant responded, \textit{``oh,
absolutely. When I’m in meetings or have to collaborate in any way, I’m always sober for those''} [P14, US], which suggests she does consider the impacts on co-workers when making substance use decisions. 



%

\vspace{2pt}
\noindent\textbf{Substance use and safety:} Some developers also consider the \textit{risk} to software users when choosing to use a psychoactive substance. Seven participants explicitly mentioned considering the safety of users should their code go into production, often contrasting between industries (e.g., game development vs. medical technology). As one participant explained, \textit{``smoking weed in my office, I think that's not a problem as long as I'm not programming anything that's carrying risk. Like if it was a self-driving car perhaps, \dots where there's a lot of liability attached to it \dots or actually physically something could happen, that might be where the line is''} [P10, US].

This concern for risk, however, was not universal: one dissented, \textit{``morally, I don't see any problem with any psychoactive substance use during coding \dots It's not like \dots driving under the influence. Coding can't really hurt anyone''} [P15, US]. 



\begin{framed}
\noindent Participants explicitly consider impacts on communication, collaboration, and software user safety when self-regulating psychoactive substance use. 
\end{framed}


\subsection{Substance Use Visibility}
\label{sec:substanceVisibility}

\noindent\textbf{Do developers disclose substance use?} Nineteen of our participants mentioned disclosing at least some of their substance use to others at work. However, the method, manner, and reception of that disclosure varies widely by substance and individual participant. 

We also asked if participants knew, or had heard of, others (e.g., co-workers or managers) using psychoactive substances in the workplace: seventeen participants reported knowing or hearing first-hand. Of the nine who had not, one had heard rumors and three had heard of use in non-programming contexts. Alcohol (12), cannabis (10) and psychoactive prescription medication (9) were the substances
participants had most heard of others using while completing software tasks. 
Psychedelics (4) and all others (1) were not as commonly encountered. 
This is important because the contrast between what may be commonly used (i.e., stimulants, see Table~\ref{tab:substancesUsed}) and what people hear about (i.e., alcohol) suggests
that open disclosure of some substances is not common in the corporate cultures of our participants. 

\vspace{2pt}
\label{sec:HappyHour} 
\noindent\textbf{Visible work use:} Fourteen participants reported observing developers use psychoactive substances together at work or work-sponsored functions with other developers. For most (9/14), the substance was alcohol at a company happy hour or later in the workday. For example, one participant, whose company handbook permits alcohol use in the office later in the afternoon, showed a picture of the company-stocked fridge where the \textit{``bottom half is just different beers and wines''} [P6, US]. Talking about the work culture at a start-up he worked for, another participant said, \textit{``there were a lot of people who drank a lot, like quite frequently. And at a lot of team events people would definitely get really drunk''} [P10, US]. Taken together, these experiences point to a culture of alcohol acceptance at many software workplaces, an acceptance that can even go further into a potentially contentious cultural belief that alcohol can even improve programming. One participant captured this tension: \textit{``It's really weird because people think that if you drink it's OK \dots a myth that people who write code can drink beer and write this code during the night and by the morning, it will be perfect. God, no. It will never be perfect code if you drink beer all night and try to write code''} [P3, US].

Though stimulants were only mentioned by two participants in this context, it is notable that both started using stimulants in software because they saw a co-worker using it to improve focus and wanted the same benefits in their own work. One participant who works at a FAANG-category company described this experience: \textit{``At least in my workplace, it's certainly not taboo to talk about Adderall\dots I actually learned about [a type of prescription stimulant] at the workplace from a friend who gave me ten strip and was like, `hey, if you're having issues [focusing], have you tried Modafinil?', and \dots so they just went into their desk and pulled me out of a blister pack of ten and said `try it some time. Try it in the morning because it'll keep you up if you try too late.' And that's all the medical advice I got''} [P9, US]. We note that in studies of other populations (e.g., college students, cf.~\cite{Benson2015MisuseOf}), misuse of stimulant medication (including sharing of prescription stimulants) is associated with a higher risk for adverse effects. Both participants in our sample went on to get prescriptions for stimulants from psychiatrists. However, this still highlights the interconnection between company productivity culture
and substance use, as well as the potential risks of policies that discourage open discussions. 

\vspace{2pt} \label{sec:remoteWork} 
\noindent\textbf{Substance use and remote work:} As most of our participants were working in software both before and during the COVID-19 pandemic, they experienced both in-person and also remote or hybrid environments. Overall, 12 responded that their substance use has \emph{increased} during the pandemic and only 1 reported a decrease. For eight of those reporting an increase, that increase was specifically cannabis or alcohol. The primary reasons reported for this were greater substance convenience and less worry about co-workers or superiors finding out. As one explained, \textit{``You can't smoke weed at an office. And even if I could, it doesn't just feel like right to do it, like go downstairs to smoke some, and then come back. That simply doesn't work''} [P21, Australia]. This is an important consideration as companies increasingly adopt post-pandemic work-from-home policies (e.g., to support neurodivergent programmers~\cite{Das2021TowardsAccessible} or in Agile contexts~\cite{Griffin2021ImplementingLean}).

\begin{framed}
\noindent Alcohol and prescription stimulants are more likely to be used and discussed than other psychoactive substances. Cannabis and psychedelics are more taboo. Both workplace productivity culture and work-from-home policies can be associated with \emph{increases} in substance use. 
\end{framed}

\section{Organizational Policy} 
\label{sec:AnalysisOrganization}

While substance use is conventionally considered a personal or cultural topic, in our interviews, the ramifications in software also include corporate drug policies. 
Beyond the impacts on remote work discussed in Section~\ref{sec:remoteWork}, 
we also analyze interactions between psychoactive substance use and organizational policy (\textbf{RQ5}, see Section~\ref{sec:study-methodology}). To do so, we first discuss participants' views of their companies' drug policies as well as the impacts of those policies on job hiring and retention. We conclude by discussing changes our participants desire for drug culture and policy in software as a whole.


\subsection{Drug Policy in Software: General Experiences}
\label{subsec:drugPolicyGen}
We first consider participants' general experiences with, and opinions on, drug policies at software work places. In our analysis, 25 of our participants spoke on 
software organization drug policies. We discuss three main sub-themes: the predominance of implicit messaging in software drug policies, participant experiences with drug tests, and the reported ineffectiveness of many software anti-drug policies.

\vspace{2pt}
\noindent\textbf{Implicit drug policies:} For the majority of participants (15/26), drug policies at their current workplaces are either primarily implicit, do not exist, or are not consistent with visible developer behavior. Developers are split on if they would prefer a more explicit policy, with some worried that being more explicit would curtail or police their substance use. However, according to several participants, implicit messaging around drug policies can lead to the necessity to navigate nuance more than desired. As explained by one participant at an office with a de facto alcohol policy that is more permissive than the official one, \textit{``there's just that tiny bit of like it could be used against me. You know. It's a lingering thought. I mean,\dots I don't think that that would be the case. But heaven forbid that there is a moment where\dots a group of us try to grab a beer from the fridge at four o'clock or 4:29, and they use that as opportunity for reprimand. Yeah. Past trauma. It's not related to current leadership, but yeah, the past''} [P6, US].

In another example of how drug policies often require nuance to interpret, one participant at a FAANG-category company explained how company policies around prescription stimulants lead to potentially unexpected cultural impacts: \textit{``So we have this health center on campus that's got doctors, nurses, lab on-site, pharmacy\dots And there are rumors about the place where if you just go in for an appointment and you talk about having focus problems, it's pretty known that they are easy for writing Adderall scripts. And then,\dots the pharmacy will waive your copay\dots The rumors about how easy it is to get an ADHD prescription and then also the implicit acknowledgment and waiving the copay if you fill it up at the company pharmacy... it sends an interesting message''} [P9, US]. 


\vspace{2pt}
\noindent\textbf{Experiences with drug tests:} The most common explicit anti-drug action in our data was drug testing: In our sample, only 38.5\% (10/26) of participants had ever taken a drug test for a software-related job. While slightly higher, given the sample size, this percentage is not significantly different from the 29\% reported by Endres \emph{et al.}~\cite{Endres2022HashingIt}. However, due to the qualitative nature of our data, we are able to elaborate with more nuance: for all but two drug-tested participants, the drug-testing was limited to an initial screening test during hiring. For the two remaining participants, one only had to be tested before driving the company van. Thus, only one participant reported regularly receiving drug tests during their software job. An additional two participants also indicated that, while there was no regular testing, there was always a threat of random drug testing should their job performance suffer.

Even though the actual number of tests taken by most participants is low, potential tests do lead to additional stress and frustration. For example, three participants indicated that the existence of a hiring drug test screening was not adequately communicated during the hiring process. As an example of this sentiment, one participant stated that this initial test \textit{``kind of snuck up on me. So I actually moved from California to New York City, started on-boarding and then they gave me the test. \dots if there was a problem with it, then that whole process of moving across the country and all of that, it would've been a huge problem''}. These experiences indicate that some software companies may benefit by being more explicit with their anti-drug policies during the hiring process itself.  

\vspace{2pt}
\noindent\textbf{Do anti-drug policies even work?} One of the most common themes expressed regarding software anti-drug policies is that they are \textit{ineffective}. Eight participants indicated that they found all or part of their current company's anti-drug regulations to be ineffective. For example, several described bypassing initial drug screening requirements through temporarily abstaining from psychoactive substance use. The ineffectiveness of anti-drug policies seems to be \textit{increased by remote work}: as one participant who works remotely for an international company states brusquely, \textit{``Honestly, like is the company in Nottingham going to come piss test me in the U.S.? No. Totally ineffective''} [P12, US]. The observation that many substance users may view extant drug policies as ineffective and easy to circumvent has significant implications on drug policies in software. If
current anti-drug policies are ineffective, companies may
benefit from reevaluating the cost-benefit trade off they
embody or more clearly communicating why they are present.
For example, participants were more understanding if
the drug test was a legal requirement the company 
could not control, \textit{``it would be a positive signal to work culture to say like, ``Hey, this, we're acknowledging that this may be a little prescriptive or archaic''. If it's for legal reasons, I think people are very understanding of it''} [P6, US]. By 
contrast, policies that lean more to ``security theater'' 
may be both a poor use of company resources and a detriment to software culture. 


\begin{framed}
    \noindent  While the number of drug tests required by a software job is typically low, poor communication surrounding initial screening tests can still influence candidate decisions. 
    At the same time, current software drug policies are often viewed as ineffective, a feeling mediated by remote work. These two results encourage revisiting the costs and benefits of anti-drug policies for software jobs.
\end{framed}

\subsection{Drug Policy Impacts: Hiring and Retention}

We also asked if a policy has or would impact the decision to work at a company. Overall, 11 out of 26 participants said an organization's policy around psychoactive substance use would or has influenced their decisions to work there. A further five indicated that it might impact their decisions, depending on how restrictive the policy was or how much they wanted that specific job. Together, for 16 out of 26 participants, a drug policy could impact job hiring or retention. As both the pre-study (see Section~\ref{sec:preStudy}) and also prior literature indicate that psychoactive substances use in software is widespread~\cite{Endres2022HashingIt}, this finding indicates that software company policy makers may want to consider carefully the ramifications of their organization's current drug policies on hiring and retention.

\vspace{2pt}
\noindent\textbf{Why drug policies may hurt hiring and retention:} We now detail the most common elaborations on this sentiment: a belief that a policy would be too restrictive on behaviour (e.g., they would fail certain types of policies), and a belief that such policies are a negative indicator of a company's culture. 

For half of participants who answered yes or maybe (8/16), responses were contingent on how restrictive the policy was or their unwillingness to modify their substance use behaviors to comply. Generally, participants were opposed to random drug testing at software jobs or policies that banned prescribed medications (e.g., anything that would force long-term changes in substance usage patterns), but were more understanding of an initial drug test when hiring. In an indicative quote, one participant stated \textit{``I wouldn't want random testing. Like I'm cool if you want to do the on hire testing, then sure, I can abstain [from cannabis] and bring my script for Adderall and be by the book''} [P9, US]. 
By contrast, three participants (two non-prescribed stimulant users and one cannabis user) expressed that even an initial drug screening would cause them not to apply to a job, noting 
unwillingness to change their substance use even in the short term (and thus believing they would fail any such test). Together, we find that more restrictive drug policies, especially those that admit the possibility of random testing, are more likely to cause substance-using programmers to not apply to work at a company. 

Some participants were also concerned by the
cultural implications of anti-drug policies. For example, one participant stated that the existence of a drug policy was \textit{``a huge deal breaker''} and that they \textit{``would not be comfortable working somewhere where they were going to \dots have that little trust in me''} [P20, US]. 
Similarly, a second developer also expressed that she thought drug policies at a software company would reflect negatively on the company culture, stating \textit{``I don't know, in 2022? \dots I feel like certain things are indicators of how old-fashioned or inflexible a company’s work culture is. And obviously that's not something I want, so I think [a drug policy] would definitely make me reconsider''} [P26, US].
Together, these results may indicate that the existence of anti-drug policies may make those developers who value trust, individuality, and progressive policies less likely to apply.



\vspace{2pt}
\noindent\textbf{Why drug policies may \textit{not} hurt hiring and retention:} While the majority of participants indicated that a company drug policy could impact their decision to work at that company, a substantial number (10/26) indicated that it would not. However, only one participant expressed a willingness to permanently change their substance use while programming to adhere to a company policy, stating that \textit{``it depends on the job \dots Say I'll get a job at Google, and Google will require this practice, then I'll quit smoking \dots I just like programming more than smoking''} [P3, US]. 

For other participants, their stances were always motivated by a belief the drug policy would not impact them, either because the substances they used would not be banned (e.g., they used only prescribed medications), they planned on keeping their substance use secret indefinitely regardless of the policy, or they thought any policy would be ineffective and thus not worth considering (see Section~\ref{subsec:drugPolicyGen}). As an indicative example, one participant who works at a startup in Silicon Valley connected their non-consideration of drug policies to remote work, stating \textit{``with remote work they can't really tell. So [a drug policy] wouldn't really affect me''} [P4, US].

Overall, these results indicate that developer ambivalence 
toward drug policies stems from believing those polices have no impact, rather than believing those policies would substantively change their behavior. 

\begin{framed}
    \noindent Over half of our participants (16/26) indicated that a drug policy has or could impact their decisions to work at a software job, primarily by how restrictive the policy is and what it indicates about company culture. Those participants who would not be influenced by a drug policy cited a belief in the ineffectiveness of the policy or a continued intention to keep their drug use secret, rather than a desire for, or agreement with, the policy itself. 
\end{framed}

\subsection{Drug Use in Software: What should change?}


\begin{table}[t]
\centering
\caption{Suggestion categories for changes to drug policy (pink) or drug culture ($\dagger$, blue) in software environments and participant counts per category. 
\label{tab:proposedChanges}}
\begin{tabular}{p{0.85\columnwidth}r}
\toprule
\textbf{ Desired Software Drug Culture or Policy Changes}                  & \# \\ \midrule
\rowcolor[HTML]{A7C7E7} 
$\dagger$ Decrease stigma around psychoactive substance use in software          & 13 \\
\rowcolor[HTML]{A7C7E7} 
$\dagger$ Be more open about psychoactive substance use in software                 & 8  \\
\rowcolor[HTML]{A7C7E7} 
$\dagger$ Increase acceptance of using psychoactive substances for mental health in software                                     & 6 \\
\rowcolor[HTML]{ffd1f3} 
Make software company anti-drug policies more consistent between substances                  & 3  \\
\rowcolor[HTML]{ffd1f3} 
Increased organizational support for recreational psychoactive substance use in software development                   & 3 \\
\rowcolor[HTML]{ffd1f3} 
Change software drug policies to consider only performance impacts (rather than which substances you use or don't use) & 3 \\
\rowcolor[HTML]{ffd1f3} 
Improve education on impacts of substance use on software development & 2  \\
\rowcolor[HTML]{ffd1f3} 
Make software company anti-drug policies less strict                      & 2  \\
\rowcolor[HTML]{ffd1f3} 
More explicit policies and safety training regarding psychoactive substance use in software                  & 1  \\
\rowcolor[HTML]{A7C7E7} 
$\dagger$ Combat drug-use impacts on software's productivity culture                           & 1  \\ \bottomrule

\end{tabular}

\end{table}

Finally, we asked participants what they think should change about drug culture or policy in software environments. Overall, 20 participants proposed at least one software-specific change they would like to see. In our analysis, we identified 10 different suggestions which are listed in Table~\ref{tab:proposedChanges}, ordered by how many participants suggested each one. In the rest of this section, we discuss these suggestions in more detail: we first present suggestions that are policy-related, followed by those for software drug culture in general. Because drug policies and drug use impact
work quality, socialization and culture, and hiring decisions, 
companies are likely to benefit from considering the feedback
of those most impacted when they develop and evolve their policies. 




\vspace{2pt}
\noindent\textbf{How software drug policies should change:} On the policy side, participants suggested several changes for software work environments. A full list of these policy changes is in Table~\ref{tab:proposedChanges}. However, here we emphasize two suggestions. First, three participants suggested that they would prefer to have policies embrace performance-based or behavior-based metrics, rather than banning or allowing specific substances in particular. As one participant notes, \textit{``there are a lot of people who behave inappropriately at work even if they're sober, and there are people who work better on substances \dots So, policies should focus on behavior and not what substances you do or don't use''} [P20, US]. 
Second, three participants suggested that companies should make their policies more consistent; generally when making this suggestion, participants pointed out that while alcohol and caffeine are very accepted by software culture (e.g., corporate happy hours, office coffee machines, etc., see Section~\ref{sec:HappyHour}), 
other substances that induce a similar level of impairment often are not. For example, one participant stated \textit{``Sometimes you can see that people are just wired on caffeine. And that's widely accepted, right? So why don't we accept if a guy says, `Hey, I want to come downstairs, smoke a joint and I'll be back and do great work?'''} [P21, Australia].


\vspace{2pt}
\noindent\textbf{How software drug culture should change:} 
In addition to corporate policy, 
participants also gave suggestions for software company drug culture as a whole. The most requested change (13 participants) would be decreasing stigma toward psychoactive substance use in software. Eight participants called for decreasing the stigma around prescribed medications. For example, one notes,
\textit{``I would normalize if someone have trouble focusing \dots software engineering or debugging process is itself a mind-intensive task \dots I think it should be normalized if someone takes stress or antidepressants''} [P6, US and Bangladesh].
Other than decreasing stigma, eight participants would like programmers to be more open about psychoactive substance use in general. 
These comments point toward a more common modern perspective of embracing
talking openly about that substance use, as well as other aspects of workforce diversity, 
as part of overall efforts to recruit and retain
the best software engineers, regardless of background. 


\begin{framed}
    \noindent Participants proposed 10 different categories of changes for software drug culture and policy. The most common cultural suggestion is to decrease the stigma regarding psychoactive substance use in software communities. On the policy side, participants suggested a range of changes from loosening anti-drug policies to even encouraging the use of recreational psychoactive substances while brainstorming or solving problems in software.
\end{framed}


\section{Threats to Validity and Limitations}
\label{sec:threats}

One potential threat to the validity of our study is the high proportion of stimulant users in our population compared to that in the pre-study (see Section~\ref{sec:preStudy}), thus potentially leading to an overemphasis on stimulant user experiences. The high proportion of stimulant users in our population may be explained in part by our recruitment methods (e.g., 
recruiting from the subreddit \texttt{r/adderall}). 
We note, however, as the focus of the archival data set used in the pre-study was on recreational rather than prescribed substances, it may also be that stimulant usage was under-reported in that data (a supposition supported by a cursory look at the archival data free-response questions). We leave it to future research to do a more in-depth exploration of stimulant usage in software.

One limitation of our design relates 
to the population considered, which includes only users of psychoactive substances. As a result, the experiences
described and themes identified may not generalize to non-users. We focus on users because drug use is often an inherently personal topic: experiences vary by what
psychoactive substances are used, individual motivations
for use~\cite{Anthenien2021CannabisOutcome}, the industry a user is working in, the size of
the company, and so on. 
The goal of this study is \emph{not} to make overarching statistical claims about the prevalence of certain substances and experiences regarding the use of them in software.
Instead, we describe the experiences of 
developers using psychoactive substances in a way that
admits conclusions at personal, interpersonal, and organizational levels. 
We leave it to future work to investigate  the experiences and opinions of non-users. 

Another limitation of our study is that our sample is biased toward US-based participants (19/26 were working in the United States at the time of the interview, and 17 had exclusively US-based professional programming experiences). This is an especially important bias to consider when contextualizing our results due to the different legal and cultural statuses of substances worldwide. For example, many of our participants use prescription stimulants which are more commonly prescribed in the United States as compared to other countries~\cite{Bachmann2017TrendsIn}. We have included the locations of participants for quotes when relevant. However, we encourage future work to investigate psychoactive substance use in broader populations of programmers to better understand which findings (cf.~\cite{Darshan2013AStudy}) are transferable to other countries and cultures not represented in our sample. 

\section{Conclusion}
From alcohol to Adderall, from debugging to soft skills, from mental health to social
stigma, from company culture to remote work, we find that psychoactive substance use
pervades almost all aspects of modern software development. In a qualitative,
thematic analysis of 26 hour-long interviews with professional programmers, we delve
into the personal experiences of software engineers who use psychoactive substances. 
At the individual level, we find that \textbf{alleviating mental health symptoms} or \textbf{desired programming enhancement} are
the primary motivations. In addition, a significant emphasis is placed on \textbf{productivity} 
(e.g., with stimulants seen as aiding debugging and cannabis 
and psychedelics aiding brainstorming). 
At the socialization level, participants describe a positive impact on \textbf{``soft'' skills},
as well as \textbf{visible use at work} for many substances (and increased use
under \textbf{work from home}). 
At the organizational level, there is widespread agreement that \textbf{anti-drug
policies are unclear and ineffective}. Such policies are viewed as indicative
of corporate culture and may have a \textbf{negative impact on hiring and retention}. 
To the best of our knowledge, this is the first qualitative study of modern software engineer
experiences with psychoactive substances, and we hope it will encourage further
transparent discussion of an important issue that impacts the health and happiness 
of many developers, as well as the productivity, culture, and hiring of organizations. 

\section*{Acknowledgements}

We acknowledge the partial support of the National Science Foundation (CCF 2211749) as well as the University of Michigan \textit{Center for Academic Innovation} and the University of Michigan \textit{Chronic Pain \& Fatigue Research Center}.
We thank Zachary Karas for his aid in the manual transcription of several of our interviews. Additionally, we also extend our thanks to those who gave feedback on initial versions of this work for their advice on the contextualized and measured phrasing of our findings on this sensitive topic.

\bibliographystyle{abbrv}
\bibliography{endres_bib_living}

\begin{thebibliography}{10}

\bibitem{Ahmed2015SoftSkills}
F.~Ahmed, L.~F. Capretz, S.~Bouktif, and P.~Campbell.
\newblock Soft skills and software development: {A} reflection from the
  software industry.
\newblock {\em CoRR}, abs/1507.06873, 2015.

\bibitem{Anthenien2021CannabisOutcome}
A.~Anthenien, M.~Prince, G.~Wallace, T.~Jenzer, and C.~Neighbors.
\newblock Cannabis outcome expectancies, cannabis use motives, and cannabis use
  among a small sample of frequent using adults.
\newblock {\em Cannabis}, 4(1):69--84, 2021.

\bibitem{Bachmann2017TrendsIn}
C.~J. Bachmann, L.~P. Wijlaars, L.~J. Kalverdijk, M.~Burcu, G.~Glaeske, C.~C.
  Schuiling-Veninga, F.~Hoffmann, L.~Aagaard, and J.~M. Zito.
\newblock Trends in adhd medication use in children and adolescents in five
  western countries, 2005--2012.
\newblock {\em European Neuropsychopharmacology}, 27(5):484--493, 2017.

\bibitem{Benson2015MisuseOf}
K.~Benson, K.~Flory, K.~L. Humphreys, and S.~S. Lee.
\newblock Misuse of stimulant medication among college students: a
  comprehensive review and meta-analysis.
\newblock {\em Clinical child and family psychology review}, 18(1):50--76,
  2015.

\bibitem{Brown2011StandardizedMeasures}
S.~A. Brown.
\newblock Standardized measures for substance use stigma.
\newblock {\em Drug and Alcohol Dependence}, 116(1):137--141, 2011.

\bibitem{Chattopadhyay2021DevelopersWho}
S.~Chattopadhyay, D.~Ford, and T.~Zimmermann.
\newblock Developers who vlog: Dismantling stereotypes through community and
  identity.
\newblock {\em Proceedings of the ACM on Human-Computer Interaction},
  5(CSCW2):1--33, 2021.

\bibitem{Darshan2013AStudy}
M.~Darshan, R.~Raman, T.~S. Rao, D.~Ram, and B.~Annigeri.
\newblock A study on professional stress, depression and alcohol use among
  {I}ndian {IT} professionals.
\newblock {\em Indian journal of psychiatry}, 55(1):63, 2013.

\bibitem{Das2021TowardsAccessible}
M.~Das, J.~Tang, K.~E. Ringland, and A.~M. Piper.
\newblock Towards accessible remote work: Understanding work-from-home
  practices of neurodivergent professionals.
\newblock {\em Proc. ACM Hum.-Comput. Interact.}, 5(CSCW1), apr 2021.

\bibitem{Endres2022HashingIt}
M.~Endres, K.~Boehnke, and W.~Weimer.
\newblock Hashing it out: A survey of programmers' cannabis usage, perception,
  and motivation.
\newblock In {\em International Conference on Software Engineering}, page
  1107–1119, 2022.

\bibitem{Ford2019HowRemote}
D.~Ford, R.~Milewicz, and A.~Serebrenik.
\newblock How remote work can foster a more inclusive environment for
  transgender developers.
\newblock In {\em 2019 IEEE/ACM 2nd International Workshop on Gender Equality
  in Software Engineering (GE)}, pages 9--12. IEEE, 2019.

\bibitem{Ford2022ATale}
D.~Ford, M.~D. Storey, T.~Zimmermann, C.~Bird, S.~Jaffe, C.~S. Maddila, J.~L.
  Butler, B.~Houck, and N.~Nagappan.
\newblock A tale of two cities: Software developers working from home during
  the {COVID-19} pandemic.
\newblock {\em {ACM} Trans. Softw. Eng. Methodol.}, 31(2):27:1--27:37, 2022.

\bibitem{Graziotin2019HappinessAnd}
D.~Graziotin and F.~Fagerholm.
\newblock {\em Happiness and the Productivity of Software Engineers}, pages
  109--124.
\newblock Apress, Berkeley, CA, 2019.

\bibitem{Graziotin2018WhatHappens}
D.~Graziotin, F.~Fagerholm, X.~Wang, and P.~Abrahamsson.
\newblock What happens when software developers are (un)happy.
\newblock {\em Journal of Systems and Software}, 140:32--47, 2018.

\bibitem{Griffin2021ImplementingLean}
L.~Griffin.
\newblock Implementing lean principles in scrum to adapt to remote work in a
  covid-19 impacted software team.
\newblock In A.~Przyby{\l}ek, J.~Miler, A.~Poth, and A.~Riel, editors, {\em
  Lean and Agile Software Development}, pages 177--184, Cham, 2021. Springer
  International Publishing.

\bibitem{Hipes2016TheStigma}
C.~Hipes, J.~Lucas, J.~C. Phelan, and R.~C. White.
\newblock The stigma of mental illness in the labor market.
\newblock {\em Social Science Research}, 56:16--25, 2016.

\bibitem{Holland2019RelativeAge}
J.~Holland and K.~Sayal.
\newblock Relative age and {ADHD} symptoms, diagnosis and medication: a
  systematic review.
\newblock {\em European Child \& Adolescent Psychiatry}, 28:1417--1429, 2019.

\bibitem{Huang2021LeavingMy}
Y.~Huang, D.~Ford, and T.~Zimmermann.
\newblock Leaving my fingerprints: Motivations and challenges of contributing
  to {OSS} for social good.
\newblock In {\em 43rd {IEEE/ACM} International Conference on Software
  Engineering, {ICSE} 2021, 22-30 May 2021}, pages 1020--1032, Madrid, Spain,
  2021. {IEEE}.

\bibitem{Jarosz2012UncorkingThe}
A.~F. Jarosz, G.~J. Colflesh, and J.~Wiley.
\newblock Uncorking the muse: Alcohol intoxication facilitates creative problem
  solving.
\newblock {\em Consciousness and Cognition}, 21(1):487--493, 2012.
\newblock Beyond the Comparator Model.

\bibitem{Johnson2022ProgramlOnline}
J.~Johnson, A.~Begel, R.~Ladner, and D.~Ford.
\newblock Program-l: Online help seeking behaviors by blind and low vision
  programmers.
\newblock In {\em 2022 IEEE Symposium on Visual Languages and Human-Centric
  Computing (VL/HCC)}, pages 1--6. IEEE Computer Society, 2022.

\bibitem{Kaur2020OptimizingFor}
H.~Kaur, A.~C. Williams, D.~McDuff, M.~Czerwinski, J.~Teevan, and S.~T. Iqbal.
\newblock Optimizing for happiness and productivity: Modeling opportune moments
  for transitions and breaks at work.
\newblock In {\em Proceedings of the 2020 CHI Conference on Human Factors in
  Computing Systems}, CHI '20, page 1–15, New York, NY, USA, 2020.
  Association for Computing Machinery.

\bibitem{Kendall2016HackingThe}
M.~Kendall.
\newblock Hacking the brain: {Silicon} {Valley} entrepreneurs turn to fasting
  and smart drugs.
\newblock {\em The Mercury News}, 9, 2016.

\bibitem{Kim2016TheEmerging}
M.~Kim, T.~Zimmermann, R.~DeLine, and A.~Begel.
\newblock The emerging role of data scientists on software development teams.
\newblock In {\em Proceedings of the 38th International Conference on Software
  Engineering}, ICSE '16, page 96–107, New York, NY, USA, 2016. Association
  for Computing Machinery.

\bibitem{Markoff2005WhatThe}
J.~Markoff.
\newblock {\em What the dormouse said: How the sixties counterculture shaped
  the personal computer industry}.
\newblock Penguin Group, New York, NY, USA, 2005.

\bibitem{Matturro2019ASystematic}
G.~Matturro, F.~Raschetti, and C.~Fontán.
\newblock A systematic mapping study on soft skills in software engineering.
\newblock {\em JUCS - Journal of Universal Computer Science}, 25(1):16--41,
  2019.

\bibitem{McDonald2019ReliabilityAnd}
N.~McDonald, S.~Schoenebeck, and A.~Forte.
\newblock Reliability and inter-rater reliability in qualitative research:
  Norms and guidelines for cscw and hci practice.
\newblock {\em Proceedings of the ACM on human-computer interaction},
  3(CSCW):1--23, 2019.

\bibitem{McIntosh2015SituatingAnd}
M.~J. McIntosh and J.~M. Morse.
\newblock Situating and constructing diversity in semi-structured interviews.
\newblock {\em Global Qualitative Nursing Research}, 2:2333393615597674, 2015.
\newblock PMID: 28462313.

\bibitem{Morris2015UnderstandingThe}
M.~R. Morris, A.~Begel, and B.~Wiedermann.
\newblock Understanding the challenges faced by neurodiverse software
  engineering employees: Towards a more inclusive and productive technical
  workforce.
\newblock In {\em SIGACCESS Conference on Computers \& Accessibility}, page
  173–184, 2015.

\bibitem{Newman2023MaterialsFor}
K.~Newman, M.~Endres, W.~Weimer, and B.~Johnson.
\newblock {Materials for From Organizations to Individuals: Psychoactive
  Substance Use By Professional Programmers}.
\newblock Zenodo, Feb. 2023.
\newblock {We note, due to the sensitivity of our data, the interview data
  itself is not included here. Please contact the authors to learn more if
  interested (weimerw@umich.edu) or (endremad@umich.edu).}

\bibitem{nida2018PrescriptionStimulants}
NIDA.
\newblock Prescription stimulants drugfacts.
\newblock
  \url{https://nida.nih.gov/publications/drugfacts/prescription-stimulants},
  2018.
\newblock Accessed: 2022-08-28.

\bibitem{Prochazkova2018ExploringThe}
L.~Prochazkova, D.~P. Lippelt, L.~S. Colzato, M.~Kuchar, Z.~Sjoerds, and
  B.~Hommel.
\newblock Exploring the effect of microdosing psychedelics on creativity in an
  open-label natural setting.
\newblock {\em Psychopharmacology}, 235(12):3401--3413, 2018.

\bibitem{Ransing2022CurrentState}
R.~Ransing, P.~A. de~la Rosa, V.~Pereira-Sanchez, J.~I. Handuleh, S.~Jerotic,
  A.~K. Gupta, R.~Karaliuniene, R.~de~Filippis, E.~Peyron,
  E.~S{\"o}nmez~G{\"u}ng{\"o}r, et~al.
\newblock Current state of cannabis use, policies, and research across sixteen
  countries: cross-country comparisons and international perspectives.
\newblock {\em Trends in psychiatry and psychotherapy}, 44, 2022.

\bibitem{Sadler2010ResearchArticle}
G.~R. Sadler, H.-C. Lee, R.~S.-H. Lim, and J.~Fullerton.
\newblock Research article: Recruitment of hard-to-reach population subgroups
  via adaptations of the snowball sampling strategy.
\newblock {\em Nursing \& Health Sciences}, 12(3):369--374, 2010.

\bibitem{Seaman2008QualitativeMethods}
C.~B. Seaman.
\newblock Qualitative methods.
\newblock In {\em Guide to advanced empirical software engineering}, pages
  35--62. Springer, 2008.

\bibitem{Singh2019WomenParticipation}
V.~Singh.
\newblock Women participation in open source software communities.
\newblock In {\em Proceedings of the 13th European Conference on Software
  Architecture-Volume 2}, pages 94--99, 2019.

\bibitem{SubstanceAbuseandMentalHealthServicesAdministration2020KeySubstance}
{Substance Abuse and Mental Health Services Administration}.
\newblock Key substance use and mental health indicators in the united states:
  results from the 2020 national survey on drug use and health.
\newblock {\em Center for Behavioral Health Statistics and Quality, Substance
  Abuse and Mental Health Services Administration}, 2020.

\bibitem{Taris2009WellbeingAnd}
T.~W. Taris and P.~J. Schreurs.
\newblock Well-being and organizational performance: An organizational-level
  test of the happy-productive worker hypothesis.
\newblock {\em Work \& Stress}, 23(2):120--136, 2009.

\bibitem{Wadley2016HowPsychoactive}
G.~Wadley.
\newblock How psychoactive drugs shape human culture: A multi-disciplinary
  perspective.
\newblock {\em Brain Research Bulletin}, 126:138--151, 2016.
\newblock Neurobiology of emerging psychoactive drugs.

\bibitem{Walsh2011DrugsThe}
C.~Walsh.
\newblock Drugs, the internet and change.
\newblock {\em Journal of psychoactive drugs}, 43(1):55--63, 2011.

\bibitem{Whiting2008SemistructuredInterviews}
L.~S. Whiting.
\newblock Semi-structured interviews: guidance for novice researchers.
\newblock {\em Nursing Standard}, 22:35--40, 2008.

\bibitem{WHO2022DrugsPsychoactive}
WHO.
\newblock Drugs (psychoactive), 2022.
\newblock Accessed: 2022-08-25.

\bibitem{Zelenski2008TheHappyproductive}
J.~Zelenski, S.~Murphy, and D.~Jenkins.
\newblock The happy-productive worker thesis revisited.
\newblock {\em J. Happiness Studies}, 9:521--537, 2008.

\end{thebibliography}

\end{document}